\documentclass[conference]{IEEEtran}
\IEEEoverridecommandlockouts
\usepackage{cite}
\usepackage{amsmath,amssymb,amsfonts,mathrsfs}
\usepackage{algorithm,algorithmic}
\usepackage{graphicx,subfig}
\usepackage{textcomp}
\usepackage{xcolor}
\usepackage{caption}
\def\BibTeX{{\rm B\kern-.05em{\sc i\kern-.025em b}\kern-.08em
    T\kern-.1667em\lower.7ex\hbox{E}\kern-.125emX}}
\begin{document}

\title{An Integrated Communication and Computing Scheme for Wi-Fi Networks based on Generative AI and Reinforcement Learning\\

\thanks{The work of X. Du, and X. Fang was supported in part by the NSFC under Grant No. 62071393. (Corresponding author: Xuming
Fang.)}
}

\author{Xinyang Du, Xuming Fang\textsuperscript{*}\\ 
Key Laboratory of Information Coding and Transmission, Southwest Jiaotong University, Chengdu, China \\ 
xydu@my.swjtu.edu.cn, xmfang@swjtu.edu.cn
}
\maketitle
\begin{abstract}
The continuous evolution of future mobile communication systems is heading towards the integration of communication and computing, with Mobile Edge Computing (MEC) emerging as a crucial means of implementing Artificial Intelligence (AI) computation. MEC could enhance the computational performance of wireless edge networks by offloading computing-intensive tasks to MEC servers. However, in edge computing scenarios, the sparse sample problem may lead to high costs of time-consuming model training. This paper proposes an MEC offloading decision and resource allocation solution that combines generative AI and deep reinforcement learning (DRL) for the communication-computing integration scenario in the 802.11ax Wi-Fi network. Initially, the optimal offloading policy is determined by the joint use of the Generative Diffusion Model (GDM) and the Twin Delayed DDPG (TD3) algorithm. Subsequently, resource allocation is accomplished by using the Hungarian algorithm. Simulation results demonstrate that the introduction of Generative AI significantly reduces model training costs, and the proposed solution exhibits significant reductions in system task processing latency and total energy consumption costs. 
\end{abstract}

\begin{IEEEkeywords}
mobile edge computing, reinforcement learning, generative artificial intelligence, computing-communication integration, offloading decision
\end{IEEEkeywords}

\section{Introduction}
With the rapid proliferation of Internet of Things (IoT) technology, latency-sensitive applications such as autonomous driving, video streaming analysis, virtual reality (AR/VR), and online gaming continue to emerge. There is an increasingly urgent demand for computing services with high speed, low latency, and low energy consumption. Meanwhile, the exponential growth of smart devices rapidly connecting to networks generates a large amount of data. Traditional cloud computing architectures are currently unable to achieve large-scale real-time computing for massive front-end devices. In order to significantly enhance the data processing of wireless edge networks and meet the demand of users for computing service quality, edge computing technology has emerged. 

Mobile Edge Computing (MEC) enhances the computational performance of wireless edge networks by offloading locally intensive tasks to MEC servers\cite{b1}. As a pivotal technology for integrating communication and computing, MEC meets critical requirements in various areas such as real-time operations, data optimization, security, and privacy protection. Hence, it finds widespread application in scenarios including vehicular networks, unmanned aerial vehicles, smart cities, and virtual reality services.

In recent years, people have devoted significant research efforts to the joint optimization of task offloading policies and resource allocation. The work in\cite{b3} proposed a distributed algorithm based on game theory, which jointly optimizes latency and energy consumption. An MEC system architecture with multiple access points (APs) and STAs was constructed based on the 802.11ac standard in\cite{b4}. The work proposes a scheme combining computing offloading and resource allocation. The optimization problem was formulated as an integer programming problem and solved using branch and bound method. However, the above studies all assume that communication resources can be arbitrarily partitioned, which is not applicable to existing Wi-Fi networks. Additionally, due to the complexity and variability of the transmission environment, the actual wireless channel exhibits time-varying characteristics. The solutions mentioned above become impractical. As a branch of artificial intelligence (AI), reinforcement learning (RL) has made significant contributions to decision-making problems in the MEC field. The work in\cite{b5} introduced a resource allocation scheme based on state-action-reward-state-action (SARSA) algorithm to solve resource management problems in MEC systems, aiming to minimize the weighted sum of energy consumption and latency. However, there is an issue of low sample efficiency in edge computing scenarios. Due to the lack of expert datasets, deep reinforcement learning (DRL) model typically requires extensive interaction with the environment, leading to high computational costs and time consumption\cite{b6}. 

To tackle this problem, the Generative Diffusion Model (GDM) was proposed in\cite{b7}. As a generative AI technique, it has the ability to capture complex data distributions and can seamlessly integrate with other RL policies to reduce the number of samples required and enhance RL performance\cite{b8}. GDM utilizes a denoising network to iteratively converge to an approximation of the true sample  through a series of estimation steps\cite{b9}. After obtaining the initial input, GDM gradually introduces Gaussian noise through a forward diffusion process. Subsequently, a neural network is trained to predict the noise and perform reverse diffusion, thereby completing the recovery of data and content. The work in\cite{b10} combines the GDM model with RL algorithms to propose the Diffusion Q-Learning algorithm. This model outperforms traditional DRL algorithms significantly and exhibits high scalability and flexibility, making it suitable for solving various optimization problems in wireless networks. However, there is currently scarce literatures utilizing deep diffusion reinforcement learning models to solve edge computing offloading problems. 

In response to the deficiencies identified in the previous studies, this paper proposes an optimized offloading decision and resource allocation solution based on generative AI and RL algorithms. The main contributions are as follows:
\begin{itemize}
\item In the scenario of communication-computing integration based on the 802.11ax Wi-Fi network, we have constructed a multi-user MEC offloading decision and resource allocation system model. Furthermore, based on the resource allocation characteristics of Wi-Fi, we propose a resource allocation scheme utilizing the Hungarian algorithm.
\item To address the issue of sparse samples, we proposes a task offloading decision solution based on generative AI and DRL algorithms, named Diffusion Twin Delayed DDPG (DTD3). The GDM is utilized as the policy network for the TD3 algorithm to solve the edge computing offloading decision problem. This approach significantly reduces convergence time while minimizing the weighted sum of latency and power consumption.
\end{itemize}

The rest of the paper is organized as follows. In Section II, we present the system model, including wireless transmission model, computing model, and formulate the optimization problem. In Section III, we describe the detailed design of the proposed solution based on generative AI and RL. Section IV elaborates on the performance evaluation of the proposed solution. Finally, we conclude the paper in Section V.
\section{System Model}

\subsection{Network Model}
\begin{figure}[htbp]
    \centering
\includegraphics[width=0.6\linewidth]{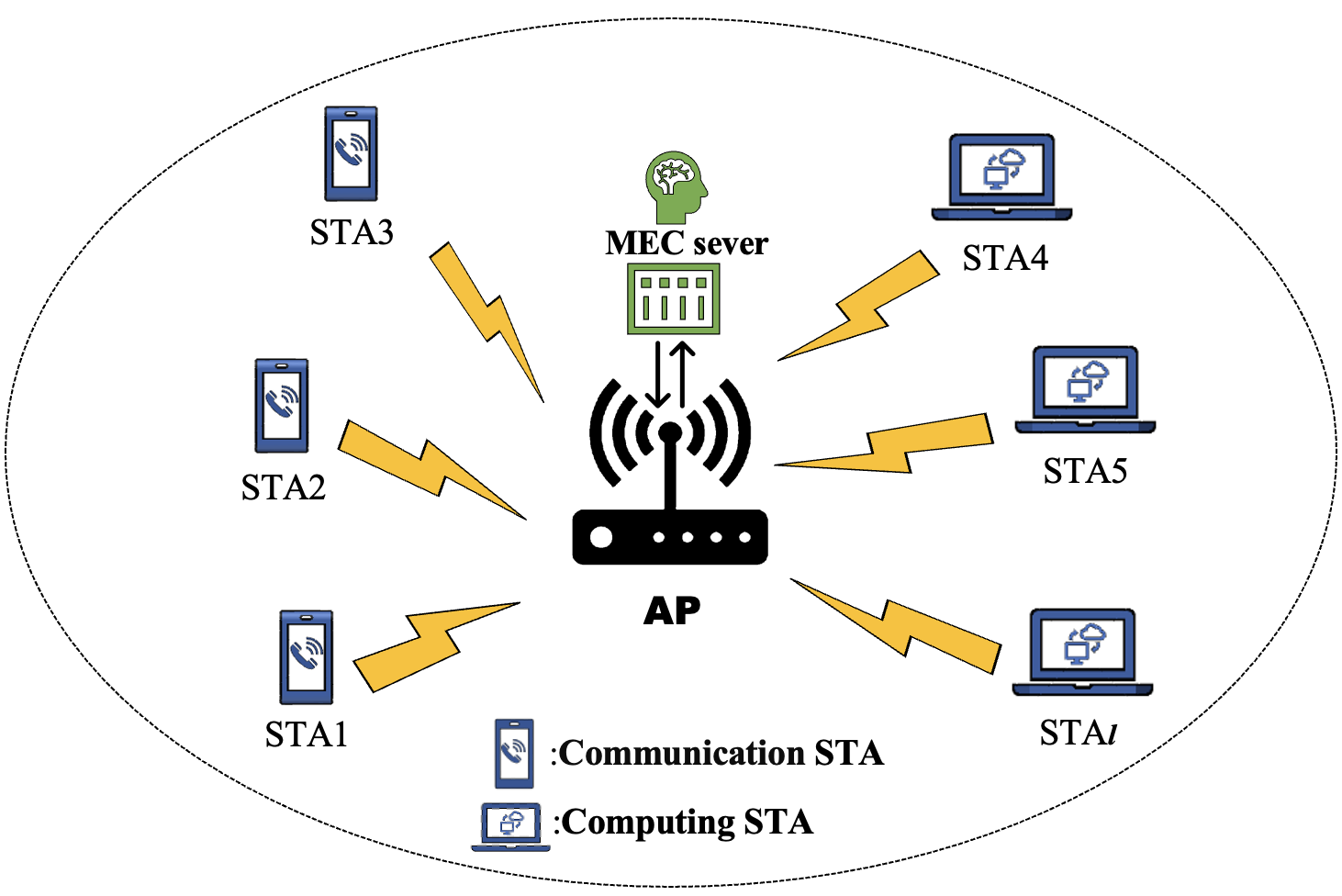}
    \caption{System model}
    \label{fig:System Model}
\end{figure}

We construct an MEC edge-end architecture based on 802.11ax Wi-Fi, as illustrated in Fig.~\ref{fig:System Model}. The scenario includes a single AP and multiple STAs. The MEC edge servers are deployed in the AP side, equipped with certain communication, computing, and storage capabilities, with the CPU computing power of MEC far exceeding that of STAs. Assuming there are a total of $L$ STAs in the scenario, denoted as $\mathscr{L}=\{1,2,3,\cdots,L\}$, among which there are $M$ computing STAs and $N$ communication STAs, denoted as $\mathscr{M}=\{1,2,3,\cdots,M\}$ and $\mathscr{N}=\{1,2,3,\cdots,N\}$ respectively. Computing STAs are limited by their own power and computing capabilities and can choose to migrate computing-intensive tasks to MEC servers for data processing to meet the computing needs of latency-sensitive tasks. Communication STAs only generate uplink traffic, requiring bandwidth from AP for data interaction, without requiring AP to allocate computing resources. 

We divide time into equal-length time slot periods, denoted as $\mathcal{T}=\{1,2,3,\cdots,T\}$. In each time slot $t$, all STAs will generate either a communication or a computing-intensive task. For computing $STA_{m}$, only one computing-intensive task is generated per time period, denoted as ${\mathcal{J}}_{m}\triangleq(d_{m},c_{m},\tau_{m})$ . $d_{m}$ represents the amount of data required to complete the task, $c_{m}$ represents the required CPU cycles for the task, and $\tau_{m}$ represents the constrained latency for computing-intensive tasks. Assuming only uplink communication traffic exists in the scenario, communication $STA_{n}$ generates one communication task per time period, denoted as ${\mathcal{J}}_{n}\triangleq(d_{n},\tau_{n})$. $d_{n}$ represents the size of the uplink transmission data, $\tau_{n}$ represents the latency constraint for uplink packet transmission. Assuming computing tasks have lower priority than communication tasks, when both computing tasks and communication tasks exist in the system simultaneously, and the resources at the AP side are not sufficient to meet all STAs' demands, the AP prioritizes fulfilling the needs of communication STAs.

\vspace{-0.38cm}  
\setlength{\belowcaptionskip}{-0.38cm}   

\subsection{Wireless Transmisson Model}

It is assumed that AP and STAs communicate and perform tasks offloading using Orthogonal Frequency-Division Multiple Access (OFDMA) in accordance with 802.11ax protocol. In 802.11ax protocol, the spectrum resources cannot be arbitrarily divided, and the smallest divisible granularity is defined as Resource Unit (RU). There is a one-to-one correspondence between STA and RU, and RU can be divided into different specifications based on the bandwidth occupied. When the channel bandwidth is 80MHz, RU specifications include several sizes such as 26-tone, 52-tone, 106-tone, 242-tone, 484-tone, and 996-tone. For computational convenience, these different RU specifications are represented as integer multiples of 26-tone, denoted as $r_{l}\in\{1,2,4,9,18,36\}$ .

In addition, assuming there are $L$ STAs needing to transmit tasks, with total bandwidth $B^{max}$ at the AP, which needs to be divided into $L$ RUs and allocated to corresponding STAs. We define $r_{l}$ as the communication resources allocated by the AP to the $STA_{l}$, where the resource vector for each STA can be represented as $\mathscr{R}=\left\{r_{1},r_{2},\ldots,r_{l}\right\}$. Let $g_{l}$ represent the channel gain, $p_{l}$ is the uplink transmission power, $N_{0}$ is the noise power spectral density, and $b_{c}$ is the bandwidth of a single 26-tone RU. Then, we can calculate the Signal-to-Noise Ratio (SNR) when transmitting on the corresponding RU as: 
\begin{equation}
    SNR_{l}=\frac{p_{l}g_{l}}{N_{0}r_{l}b_{c}}
\end{equation}

Based on the mapping between the receiver Received Signal Strength Indicator (RSSI) and Modulation and Coding Scheme (MCS)  given in the protocol, we can calculate the noise power and derive the relationship between SNR and MCS. Therefore, based on $SNR_{l}$, the size of allocated RU specifications $r_{l}$ , the mapping function between RU and transmission rate and the mapping function between SNR and the maximum selectable MCS, the maximum achievable uplink transmission rate of $STA_{l}$ can be expressed as: 
\begin{equation}
    V_{l}=\mathrm{V}(r_{l},\mathrm{MCS}(SNR_{l}))
\end{equation}

\subsection{Computing Model}

There are two modes of task execution: local computing and offloading computing. In this paper, it is assumed that computing tasks cannot be divided and can only be offloaded as a whole. When computing locally, it is assumed that the CPU computing capacity of the local device is $f_{local}$ , and the computing demand of the $STA_{m}$ is $c_{m}$. Therefore, the delay and energy consumption of local computing generated by $STA_{m}$ can be respectively represented as:
\begin{equation}
    T_{m}^{local}=\frac{c_{m}}{f_{local}}
\end{equation}
\begin{equation}
    E_{m}^{local}= 10^{-27}\times(f_{local})^{2}\times c_{m}
\end{equation}

During offloaded computing, there are three steps: 1) the STA uploads the task; 2) the AP performs the computing; and 3) the AP transmits the result back. Since the size of the result data is much smaller compared to the task data\cite{b11}, the delay and energy consumption of this part are negligible. Assuming that the AP allocates computing resources $f_{m}$ to $STA_m$, the data transmission rate of $STA_{m}$ is $V_{m}$, the amount of data for the task is $d_{m}$, and the transmission power of $STA_{m}$ is $p_{m}$, the delay and energy consumption of offloaded computation generated by $STA_{m}$ can be expressed as: 
\begin{equation}
    T_{m}^{off}=\frac{c_{m}}{f_{m}}+\frac{d_{m}}{V_{m}}
\end{equation}
\begin{equation}
    E_{m}^{off}=p_{m}\frac{d_{m}}{V_{m}}
\end{equation}

Similarly, the delay and energy consumption generated by communication $STA_{n}$ can be expressed as follows:
\begin{equation}
    T_{n}^{trans}=\frac{d_{n}}{V_{n}}
\end{equation}
\begin{equation}
    E_{n}^{trans}=p_{n}\frac{d_{n}}{V_{n}}
\end{equation}

The calculation of total delay and energy consumption can be summarized as follows:
\begin{equation}
    T_{total}=\sum_{m=1}^{M}a_{m}T_{m}^{off}+(1-a_{m})T_{m}^{local}+\sum_{n=1}^{N}T_{n}^{trans}
\end{equation}
\begin{equation}
    E_{total}=\sum_{n=1}^{M}a_{m}E_{m}^{off}+(1-a_{m})E_{m}^{local}+\sum_{n=1}^{N}E_{n}^{trans}
\end{equation}

\subsection{Problem Formulation}

Let the decision variable of $STA_{m}$ be $a_{m}\in\{0,1\}$, where $a_{m}$ = 0 represents local execution and $a_{m}$ = 1 represents offloaded execution. Thus, the offloading decision vector can be represented as $\mathscr{A}=\{a_{1},a_{2},a_{3},\cdots,a_{M}\}$. The AP communication resource allocation vector is denoted as ${\mathscr{R}}=\left\{r_{1},r_{2},\cdots,r_{L}\right\}$, and the computation resource allocation vector is denoted as $\mathscr{F}=\{f_{1},f_{2},f_{3},\cdots f_{M}\}$. Consistent with \cite{b12} and \cite{b122}, the objective of this paper is to minimize the weighted sum of system delay and energy consumption. Therefore, the optimization problem can be expressed as:
\begin{equation}
\begin{array}{l}
\text { \textbf{P}: } \min _{\mathscr{A} \mathscr{R} \mathscr{F}}
\left(\lambda T_{total}+(1-\lambda) E_{total}\right)\\
\begin{array}{l}
\text { s.t. } 
C 1:\left(1-a_{m}\right) T_{m}^{local}+a_{m} T_{m}^{o f f} 
\leq \tau_{m}, 
\forall m \in \mathscr{M} \\
C 2: T_{n}^{trans} 
\leq \tau_{n}, 
\forall n \in \mathscr{N} \\
C 3: a_{m} \in\{0,1\}, 
\forall m \in \mathscr{M} \\
C 4: \sum_{l \in \mathscr{L}} r_{l} b_{c} \leq B^{max} \\
C 5: \boldsymbol{r}_{l} \in\{1,2,4,9,18,36\}, \forall l \in \mathscr{L} \\
C 6: \sum_{m \in \mathscr{M}} f_{m} \leq F^{max} \\
\end{array}
\end{array}
\end{equation}
where $B^{max}$ and $F^{max}$ respectively denote the maximum communication and computational resources at the AP side, while $\lambda$ represents the weighting coefficient of the objective. $C1$ indicates the constraint on the maximum delay for computational tasks, $C2$ represents the constraint on the maximum delay for communication tasks, $C3$ ensures that the offloading decision is a binary variable, $C4$ represents the constraint on AP communication resources, $C5$ indicates the constraint on the allocation of RU specifications, and $C6$ denotes the constraint on AP computation resources.

The optimization problem $\textbf{P}$ involves multiple variables, and the presence of integer variables $a_{m}$ and $r_{l}$ makes it a non-convex integer programming problem\cite{b123}, which has been proven to be NP-Hard in\cite{b13}, making it difficult to solve using traditional optimization algorithms. Therefore, we formulate $\textbf{P}$ as a Markov Decision Processes (MDP) and utilize generative AI and DRL to find the optimal values for offloading and resource allocation.

\section{Offloading and Resource Allocation Scheme Based on DTD3}

\subsection{Three Key Elements for RL }
The key to solving RL problems lies in constructing the agent and environment, which involves designing parameters such as state, action, and reward.
\begin{itemize}
\item State: We define the system state space as $\mathcal{S}=\left\{d_{1},d_{2},d_{3},\cdots,d_{L},c_{1},c_{2},c_{3},\cdots c_{M},f_{mec}\right\}$, which consists of the size of the task data, the number of CPU cycles required for computing, and the available computing resources in MEC.
\item Action: Considering that the offloading policy in the system is binary, we define the action space as $\mathcal{A}=\{a_{1},a_{2},a_{3},\cdots,a_{M}\}$ , where $a_{m}\in\{0,1\}$. 
\item Reward: Considering minimizing the weighted sum of latency and energy consumption while prioritizing communication tasks to meet latency constraints, we assume    $c_{local}$ represents the total cost of local computing,  $c_{total}$ denotes the total cost of computing  in the current time slot, $t_{n}^{trans}$ stands for the transmission time of communication $STA_{n}$, and $\tau_{n}$ represents the latency constraint of communication task. Then we can define the reward function as:

\begin{equation}
\left.\mathcal{R}=\left\{\begin{array}{cc}{{\frac{c_{local}-c_{total}}{c_{local}} }}&{,\forall t_{n}^{trans}\leq\tau_{n}}\\{-1}&{,\exists t_{n}^{trans}>\tau_{n}}\end{array}\right.\right.
\end{equation}

\end{itemize}

\subsection{DTD3 Offloading Decision Algorithm}
Considering the high training cost and low efficiency of traditional RL algorithms in edge computing scenarios, this paper proposes a deep diffusion learning model called DTD3 to jointly solve the offloading decision problem in multi-user edge networks. The algorithm architecture of DTD3 consists of several components, including: policy network, twin critic networks, target policy network, target critic network, and experience replay buffer. Unlike traditional RL algorithms, DTD3 uses a Diffusion Policy (DP) based on a diffusion model as the policy network. Similar to\cite{b10}, we define $k=\{1,2,3,\cdots,K\}$ as diffusion timestep, and the RL policy of this solution can be represented as the inverse process of conditional diffusion modeling:
\begin{equation}   
\pi_{\theta }(a|s)=\mathcal{N}(a^{K};0,I)\prod_{k=1}^{K}p_{\theta}(a^{k-1}|a^{k},s) 
\end{equation}

The final sample of the reverse diffusion chain $a^0$ is the action provided by the model. Through the reverse diffusion process, the model can effectively capture the dependency between the state and the action. Generally, $p_{\theta}(a^{k-1}|a^{k},s)$ can be modeled as a Gaussian distribution $\begin{aligned}\mathcal{N}({a}^{k-1};{\mu}_{\theta}({a}^{k},{s},k),{\Sigma}_{\theta}({a}^{k},{s},k))\end{aligned}$ and can be parameterized as a noise prediction model, where $\alpha_k=1-\beta_k$, $\bar{\alpha}_k=\prod_{i=1}^k\alpha_i$, the covariance matrix is constructed as $\sum_{\theta}(a^{k},s,k)=\beta_{k}I$ and mean is represented as:
\begin{equation}    \mu_{\theta}(a^{k},s,k)=\frac{1}{\sqrt{\alpha_{k}}}\Bigg(a^{k}-\frac{\beta_{k}}{\sqrt{1-\overline{\alpha}_{k}}}\varepsilon _{\theta}(a^{k},s,k)\Bigg)
\end{equation}

The reverse diffusion process first samples $a^{K}\sim\mathcal{N}(0,I)$, and then multiple steps of sampling for $k=K,...,1$ are performed using the reverse diffusion chain parameterized by $\theta$: 
\begin{equation}    
a^{k-1}|a^{k}=\frac{a^{k}}{\sqrt{\alpha_{k}}}-\frac{\beta_{k}}{\sqrt{\alpha_{k}(1-\overline{\alpha}_{k})}}\varepsilon _{\theta}(a^{k},s,k)+\sqrt{\beta_{k}}\varepsilon 
\end{equation}
where $\varepsilon\sim\mathcal{N}(0,I)$. According to\cite{b14}, this approach defines the diffusion model loss as:
\begin{equation}
    \mathcal{L}_{d}(\theta)=\mathbb{E}_{\varepsilon\sim\mathcal{N}(0,I)}\bigg[||\varepsilon-\varepsilon_{\theta}\bigg(\sqrt{\overline{\alpha}_{k}}a^0+\sqrt{1-\overline{\alpha}_{k}}\varepsilon,s,k\bigg)||^2\bigg]
\end{equation}

Consistent with\cite{b10}, the Q-value function is injected into the reverse diffusion chain during training model:
\begin{equation}
    \mathcal{L}_{q}(\theta)=-\frac{\eta}{\mathbb{E}_{(s,a)\sim D}\left[|Q_{\phi}(s,a)|\right]}\cdot\mathbb{E}_{s\sim D,a^{0}\sim\pi_{\theta}}[Q_{\phi}(s,a^{0})]
\end{equation}
where $\eta $ is a hyperparameter used to balance the regularization and the ability of Q-learning. Thus, the final policy function learning objective can be represented as:
\begin{equation}
    \pi=\arg_{\pi_{\theta}}\mathrm{min}\bigl(\mathcal{L}_{d}(\theta)+\mathcal{L}_{q}(\theta)\bigr)
\end{equation}

The training process of DTD3 is presented in algorithm 1.
\begin{algorithm}
\caption{The training process of Diffsion TD3}
\begin{algorithmic}
\STATE {Initialize policy network $\pi_{\theta}$, twin critic networks $Q_{{\phi}_{1}}$, $Q_{{\phi}_{2}}$ and target network$\pi_{\theta}^{\prime}$, $Q_{{\phi}_{1}}^{\prime}$, $Q_{{\phi}_{2}}^{\prime}$ }            
\FOR {$t=1,2,...,T$}          
\STATE Select action by ${a}_{t}^{}\sim\pi_{\theta^{}}({a}_{t}\mid{s}_{t})$\\
\STATE Observe $r_t$,$s_{t+1}$ and store transition $(s_{t},a_{t},r_{t},s_{t+1})$ \\
\STATE Sample mini-batch $ \mathscr{B}=(s_{t},a_t,r_{t},s_{t+1})\sim \mathscr{D}$
\STATE $y\gets r(s_{t},a_{t})+\gamma\mathrm{min}_{i=1,2}Q_{\phi_{i}^{\prime}}(s_{t+1},a_{t+1}^{\prime})$

\STATE Update $Q_{{\phi}_{1}}$and $Q_{{\phi}_{2}}$ $\gets\mathbb{E}_{a_{t}^{\prime}\sim\pi_{\theta^{\prime}}}\left[||y-Q_{\phi_{i}}(s_{t},a_{t}^{\prime})||^{2}\right]$ 
\STATE Update policy network according to (18)
\IF {$t$ mod $d$}                        
\STATE Soft update target network: $\theta^{\prime}\leftarrow\tau\theta+(1-\tau)\theta^{\prime}$
\STATE $\phi_{i}^{\prime}\gets \tau\phi_{i}+(1-\tau)\phi_{i}^{\prime},\mathrm{for~\mathit{i}\in\{1,2\}}$

\ENDIF

\ENDFOR

\end{algorithmic}
\end{algorithm}

\subsection{Resource Allocation Scheme}
After completing the offloading decision, the system proceeds with the allocation of computational resources to the computing STAs. We takes into account the factors such as the required CPU computational resources $c_m$, task data size $d_m$, task latency constraint $\tau_m$ and the transmission capacity of $STA_{m}$ $capability_m$ to calculate the priority of  tasks. When an STA has poorer channel condition, larger task data size, greater CPU computational resource requirement, or stricter latency constraint, the task demand is more challenging to fulfill. Therefore, the system should assign a higher priority to such STAs and allocate more computational resources to them. We set $RU_{m}^{req}=\frac{d_{m}}{\tau_{m}}$, $f_{m}^{req}=\frac{c_{m}}{\tau_{m}}$. $capability_m$ represents the maximum MCS achievable under the current environment's channel conditions. Since the communication resources have not been allocated yet, we set $r_m$ for all STAs to 1, which means the RU specification is 26-tone. We let $w_1$, $w_2$ and $w_3$ be weight parameters, and $\sum_{i=1}^{3}w_{i}=1$. Thus, the priority calculation can be expressed as follows:
\begin{equation}
\begin{array}{c}
priority_{m} = w_1\frac{RU_{m}^{req}}{\sum_{m\in\mathscr{M}}RU_{m}^{req}}    +w_2\frac{f_{m}^{req}}{\sum_{m\in\mathscr{M}}f_{m}^{req}}\\    +w_3\frac{\frac{1}{capability_{m}}}{\sum_{m\in\mathscr{M}}\frac{1}{capability_{m}}}\end{array}
\end{equation}

The computational resource allocation is represented as:
\begin{equation}
    f_{m}=\frac{priority_{m}}{\sum_{m=1}^{M}priority_{m}}\times f_{mec}
\end{equation}

Based on the resource allocation characteristics of 802.11ax OFDMA, we consider the allocation problem of communication resources as an allocation problem of RU specifications. To address this issue, this paper first determines the combination of RU specifications used based on the number of transmission tasks of the AP. If the number of transmission tasks of the AP exceeds the maximum number of RUs that the AP can allocate, the tasks are sorted based on the calculated task priorities. RUs are allocated to tasks with higher priority, and resource redistribution is performed according to the computing resource allocation scheme. Tasks with lower priority are no longer offloaded. Given the limited bandwidth resources, the number of partitioning methods for RU specifications is limited. First, we obtain possible RU specification combinations based on the number of STAs and available bandwidth. Then, we calculate the efficiency matrix for each RU configuration combination and use the Hungarian algorithm to determine the optimal RU allocation scheme along with the corresponding STA transmission latency and energy consumption. Finally, we select the RU configuration combination that minimizes STA transmission latency and energy consumption as the final RU allocation result.
\section{Simulation Result}
We use Python 3.7 and PyTorch to build the simulation platform and conduct algorithm training and simulation. The algorithm model is deployed on the AP for centralized training in this paper. Assuming the AP's position keeps fixed, STAs are randomly distributed within a circle around the AP with the radius of 20m. All STAs complete computing tasks or communication tasks transmission via OFDMA. We modeled indoor path loss using the Keenan-Motley model. The computing task latency limit for STAs is set to 80\%-120\% of the latency when the task is computed locally. The computational capacities of MEC and STA are 10GHz and 1GHz, respectively. The transmission power of STAs is 500mW. The number of CPU cycles required for computing is uniformly distributed between 900 Megacycles and 1100 Megacycles\cite{b13}. The data size of computing task is uniformly distributed between 2.4Mbits and 4Mbits, and the communication task is between 10Mbits and 20Mbits. $\lambda $ is set to 0.8. 

To validate the superiority of our algorithm, we compares the performance of the proposed DTD3 scheme with four baseline schemes through simulations. “Local computing” means that all computing tasks are executed locally, “Full offloading” stands for that all computing tasks are offloaded, and “Random offloading” stands for that AP randomly determines the offloading decision. The above three baselines allocate RUs evenly among STAs. Further, “DQN” stands for that the offloading decision is determined by DQN model, and the RU is allocated using the Hungarian algorithm. The Quality of Service (QoS) represents the ratio of computing tasks satisfying the latency limit to the total number of computing tasks, while the communication success rate represents the ratio of communication tasks completed within the latency limit to the total number of communication tasks. We vary some parameters in simulation to observe the performance.
\subsection{Scenario with Varying Computing STAs}
\begin{figure*}[htbp]    
  \centering            
  \subfloat[]   
  {
      \label{fig:subfig1}\includegraphics[width=0.3\textwidth,height=0.14\textheight]{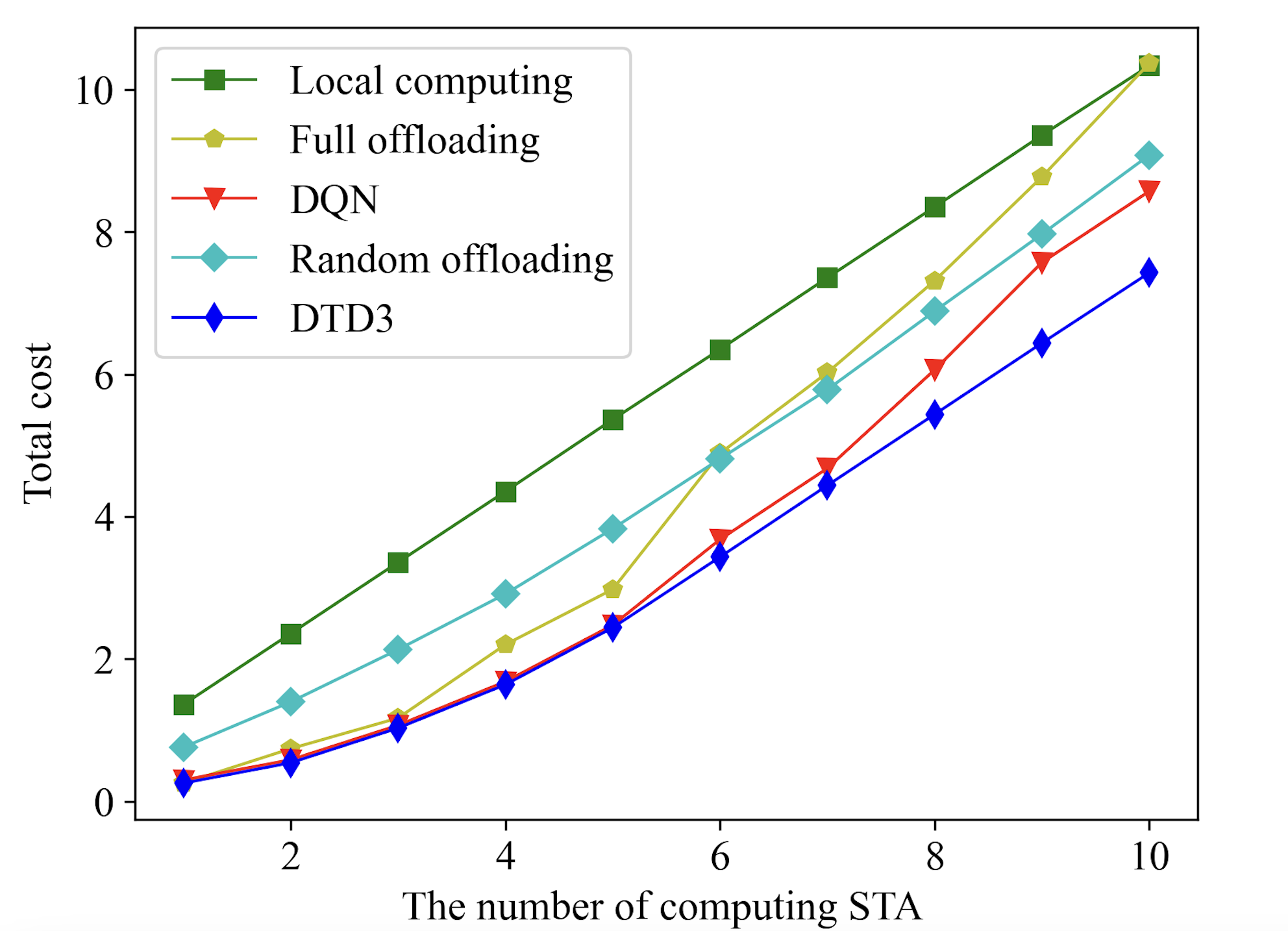}
  }
  \subfloat[]
  {
      \label{fig:subfig2}\includegraphics[width=0.3\textwidth,height=0.14\textheight]{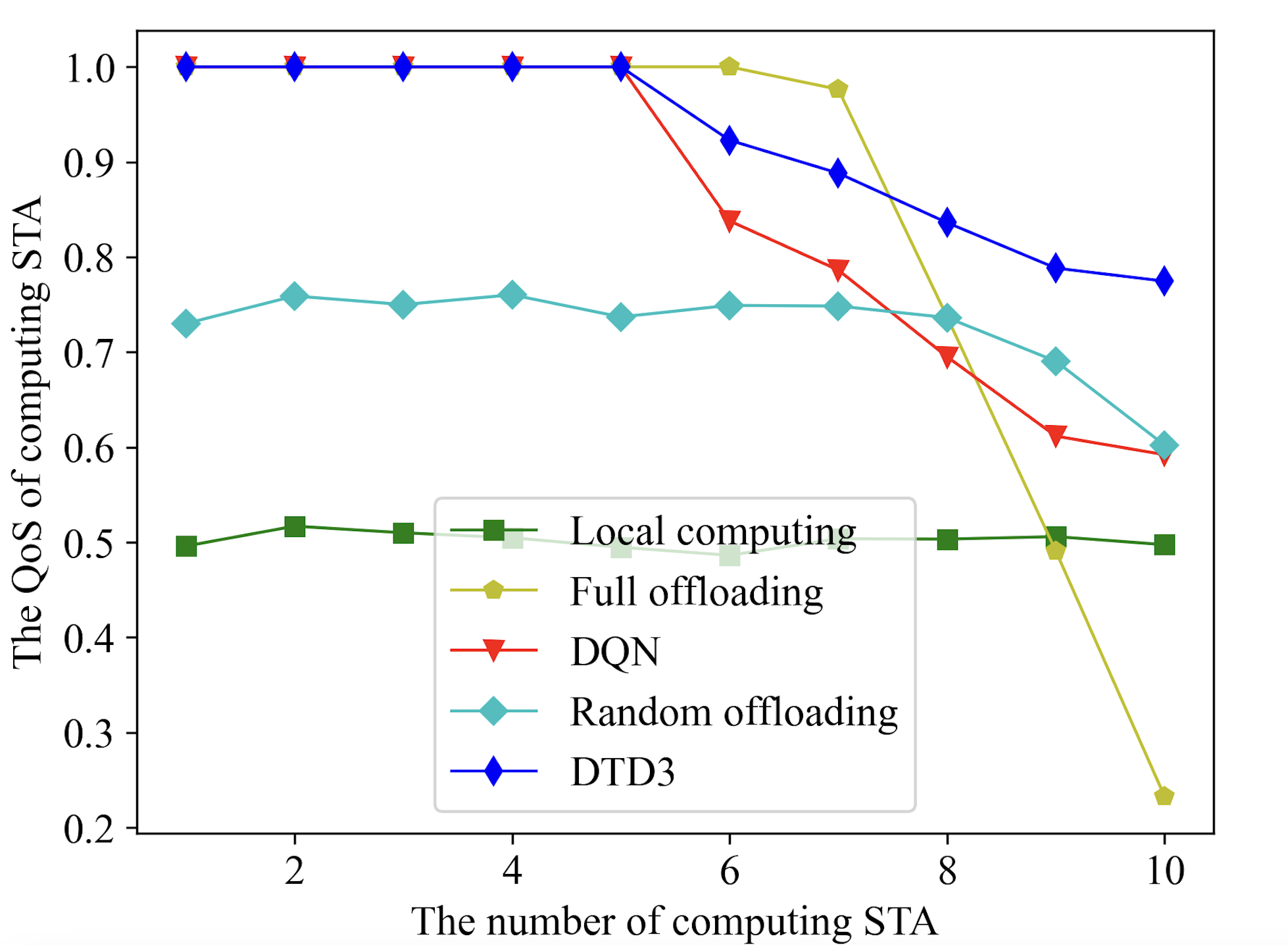}
  }
  \subfloat[]
  {
      \label{fig:subfig3}\includegraphics[width=0.3\textwidth,height=0.14\textheight]{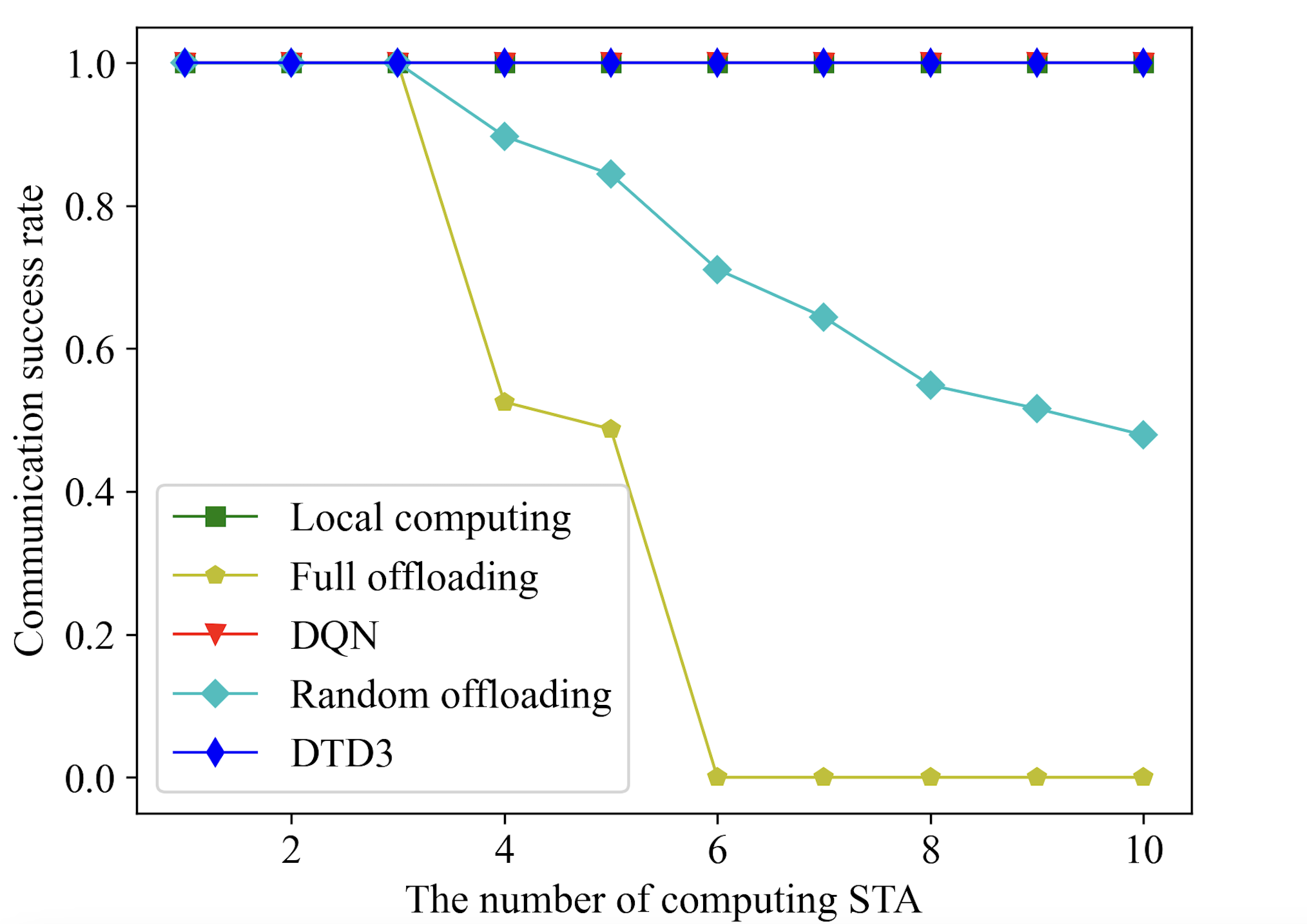}
  }
  \caption{The performance of algorithms with varying computing STAs: (a) total cost verus the number of computing STAs, (b) QoS verus the number of computing STAs, (c) communication success rate verus the number of computing STAs}    
  \label{fig:subfig_1}            
\end{figure*}

Fig.~\ref{fig:subfig_1} illustrates the performance of algorithms with varying computing STAs when there are 3 communication STAs. In Fig.~\ref{fig:subfig1}, the total cost increases with the increase in computing STAs. DTD3 demonstrates outstanding performance among all approaches. Fig.~\ref{fig:subfig2} indicates that due to the limited computing and communication resources at the AP, the QoS decreases as the number of computing STAs increases. However, under resource constraints, the QoS of DTD3 is significantly higher than other approaches. Fig.~\ref{fig:subfig3} shows that even with an increasing number of computing STAs, the communication success rate of DTD3 remains at a high level. Similarly, the trained DQN model can also provide offloading decisions that meet the communication STAs' requirements. In contrast, as the resources of the AP cannot meet the demands of all STAs, the communication success rate of the full offloading scheme significantly decreases with the increase of computing STAs.

\subsection{Scenario with Varying Capacity of MEC}

\begin{figure*}[htbp]    
  \centering            
  \subfloat[]   
  {
      \label{fig:subfig4}\includegraphics[width=0.3\textwidth,height=0.15\textheight]{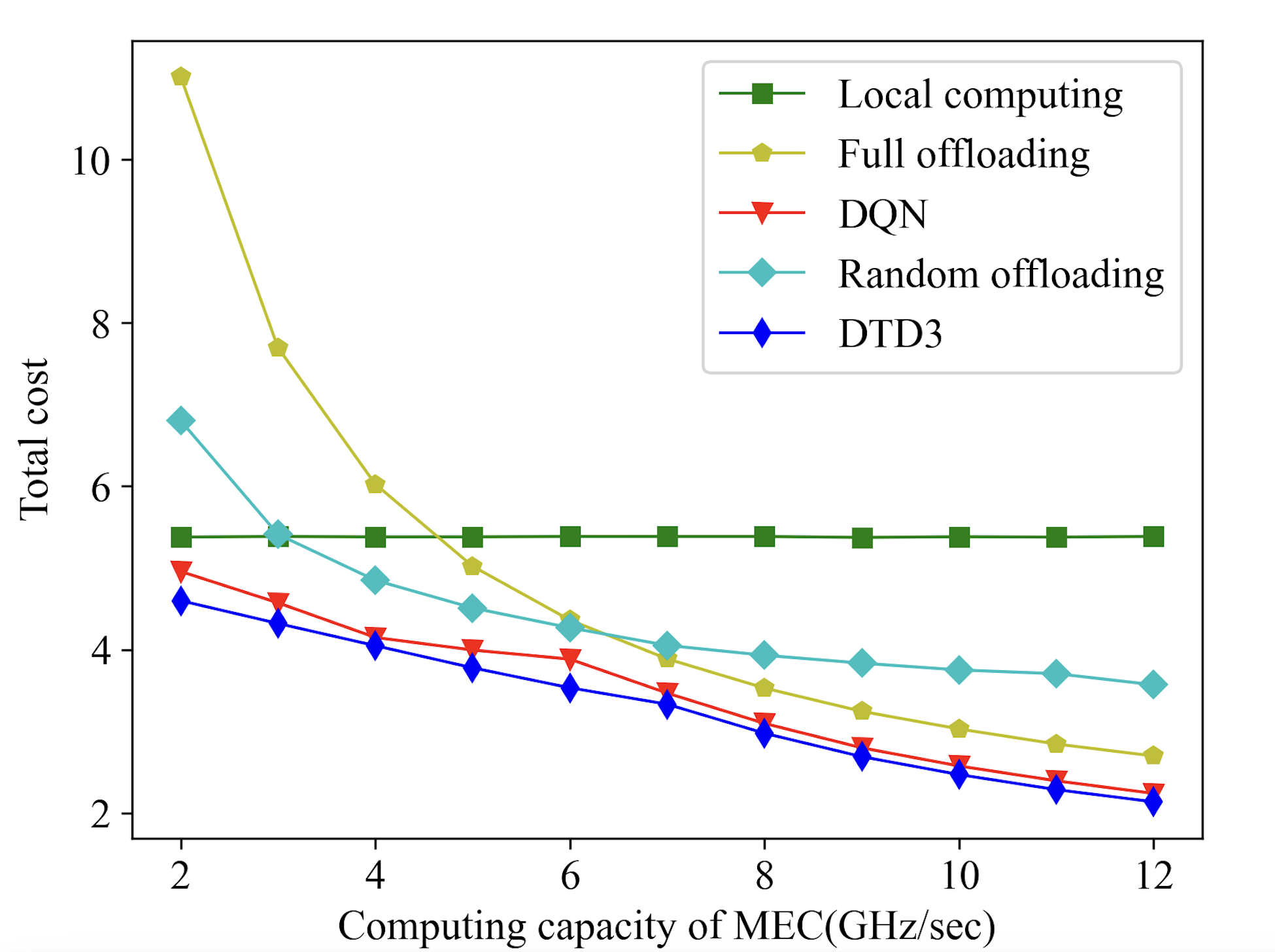}
  }
  \subfloat[]
  {
      \label{fig:subfig5}\includegraphics[width=0.3\textwidth,height=0.15\textheight]{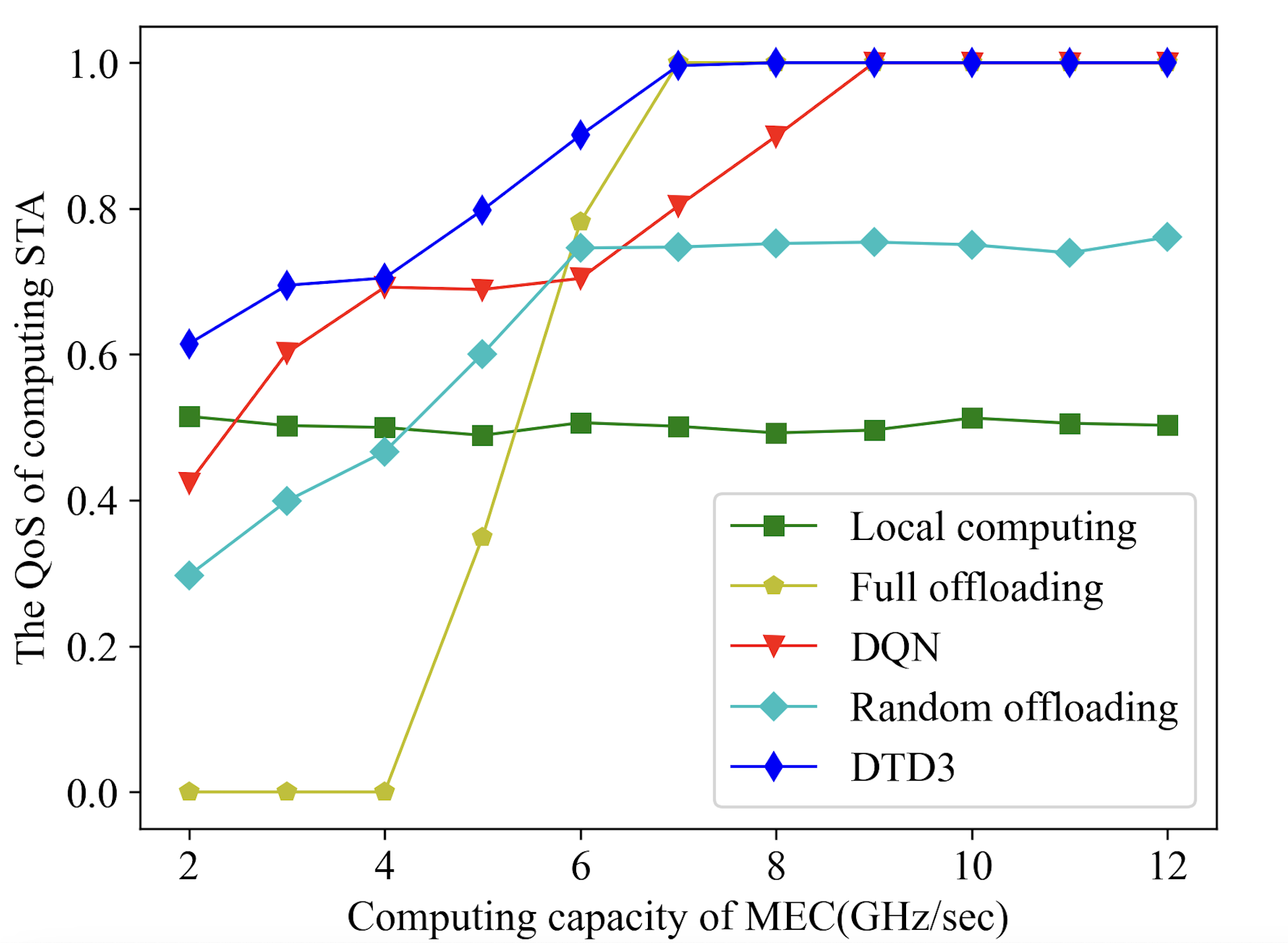}
  }
  \subfloat[]
  {
      \label{fig:subfig6}\includegraphics[width=0.3\textwidth,height=0.15\textheight]{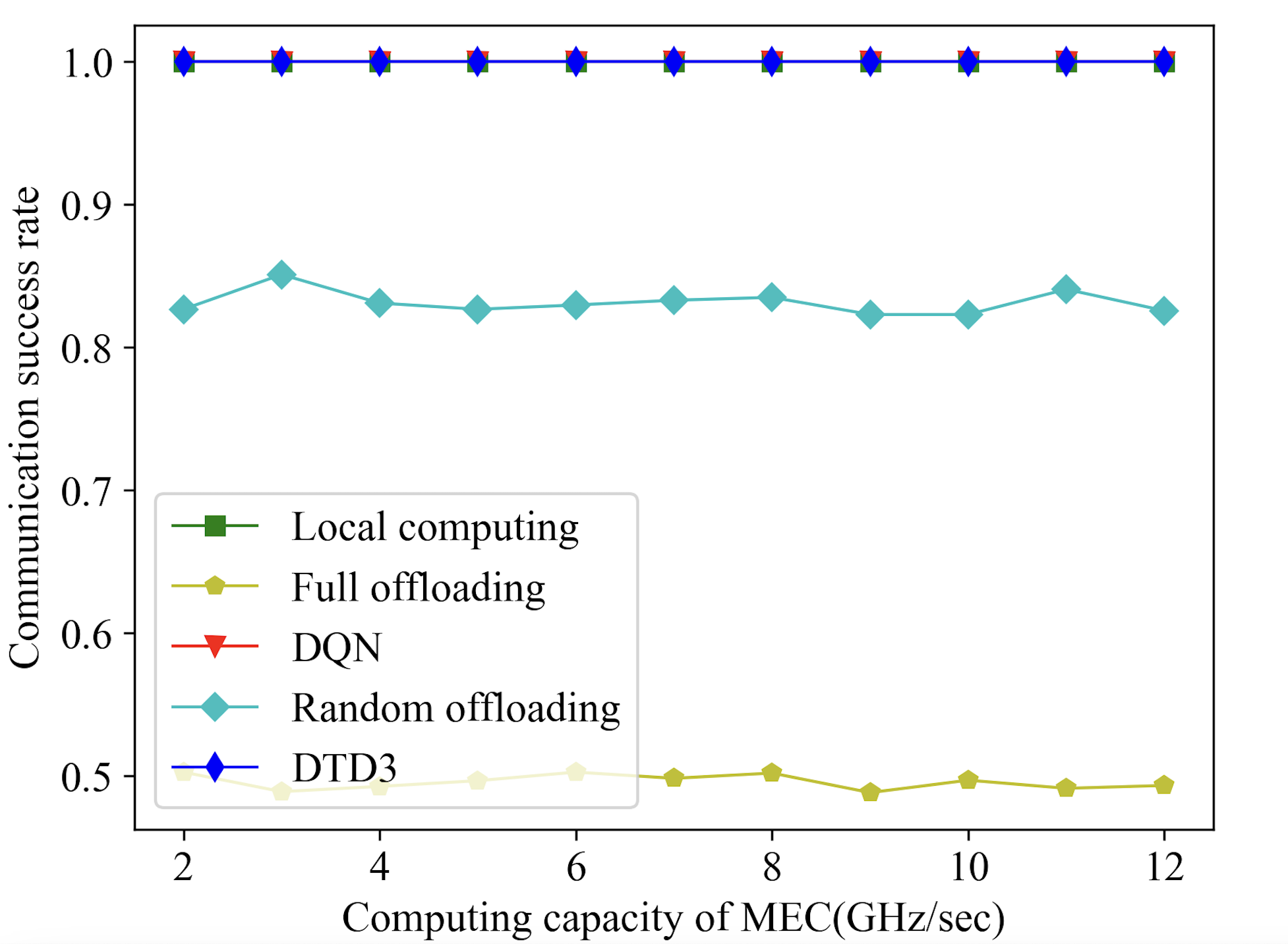}
  }
  \caption{The performance of algorithms with varying capacity of MEC: (a) total cost verus the capacity of MEC, (b) QoS verus the capacity of MEC, (c) communication success rate verus the capacity of MEC}    
  \label{fig:subfig_2}            
\end{figure*}

Fig.~\ref{fig:subfig_2}  depicts the performance of algorithms with varying capacity of MEC when there are 3 communication STAs and 5 computing STAs. Fig.~\ref{fig:subfig4} shows that as the MEC computing resource increases, the total cost decreases due to the corresponding reduction in system latency and energy consumption. In contrast, the performance trend of the local computing policy remains relatively stable as it is independent of MEC computing resources. DTD3 significantly outperforms other approaches. Fig.~\ref{fig:subfig5} illustrates that the QoS of computing STAs increases with the increase of the MEC computing resource. DTD3 achieves higher QoS compared to other approaches, particularly when MEC computing resources are limited. Fig.~\ref{fig:subfig6} shows that the communication success rate of the full offloading policy maintains around 50\%, while the random offloading policy maintains around 85\%. In contrast, the communication success rate of DTD3 and the DQN algorithm consistently maintains at 100\%.

\subsection{Convergence Analysis}
\begin{figure}[htbp]
\centering
\includegraphics[width=0.6\linewidth]{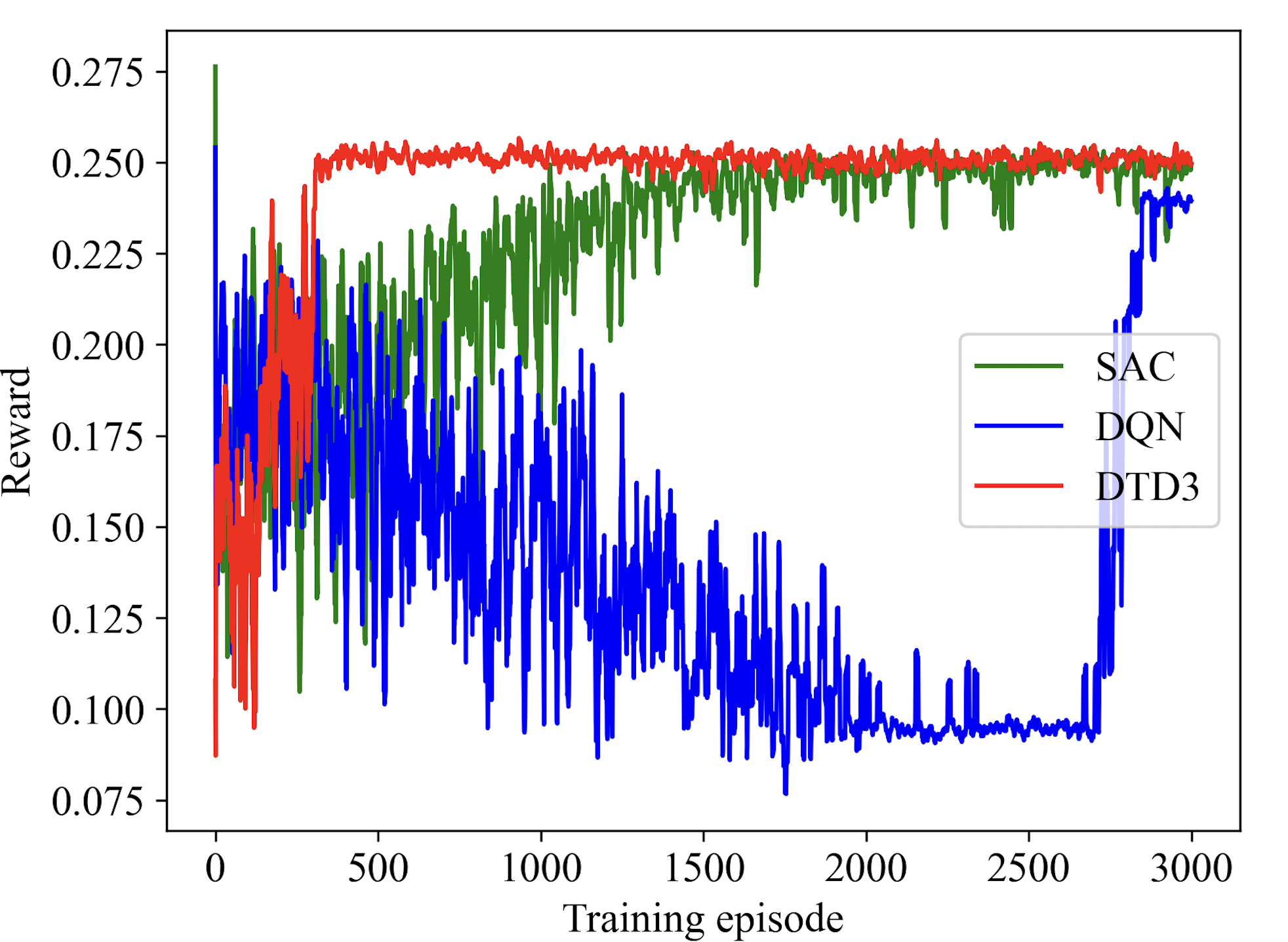}
\caption{The convergence of algorithms}
\label{fig:reward}
\end{figure}

Fig.~\ref{fig:reward}  shows the convergence performance of DQN, SAC and the proposed DTD3 algorithms when there are 3 communication STAs, 10 computing STAs, and 10GHz of MEC resources. The DTD3 algorithm converges around 400 episodes with optimal convergence performance. This is attributed to the introduced generative diffusion model, which effectively reduces the convergence time and training cost by working in coordination with the RL framework. In contrast, the convergence and stability of the DQN algorithm are relatively poor. The SAC algorithm, on the other hand, exhibits better convergence performance than the DQN algorithm due to the introduction of the maximum entropy mechanism, which enhances SAC's exploration capability and robustness.

\section{Conclusion}
In this paper, we proposes an offloading decision and resource allocation scheme based on generative AI and DRL for the integration of communication and computing in multi-STAs single-AP scenarios under 802.11ax Wi-Fi networks. We introduce diffusion models to address the sparse sample problem in edge computing scenarios and propose a communication allocation scheme more suitable for Wi-Fi environments based on the Hungarian algorithm. Extensive simulation results demonstrate that the proposed approach can reduce the overall energy consumption and latency of the system, enhance QoS, and ensure communication success rate. Moreover, compared to traditional RL methods, this model exhibits superior convergence performance. In the future, our framework could be extended to multi-agent mobile edge systems, utilizing distributed execution of RL algorithms to improve system scalability and robustness.

\end{document}